\documentclass[letterpaper]{article} 
\usepackage[margin=1.5in]{geometry}
\usepackage[hyphens]{url}  
\usepackage{graphicx} 
\usepackage[round]{natbib}
\usepackage{amsmath} 
\usepackage{amssymb}
\usepackage{booktabs}

\title{Characteristics and prevalence of fake social media profiles with AI-generated faces}
\author{
    Kai-Cheng Yang\thanks{Email: yang3kc@gmail.com} \textsuperscript{\rm 1, \rm 2}, Danishjeet Singh\textsuperscript{\rm 1}, and Filippo Menczer\textsuperscript{\rm 1} \\
    \\
    \textsuperscript{\rm 1}Observatory on Social Media, Indiana University, Bloomington, USA\\
    \textsuperscript{\rm 2}Network Science Institute, Northeastern University, Boston, USA\\
}

\newcommand{\account}[1]{{\texttt{@#1}}}
\newcommand{\package}[1]{{\texttt{#1}}}
\newcommand{\dataset}[1]{{\textbf{\texttt{#1}}}}

\begin{document}

\maketitle

\begin{abstract}
Recent advancements in generative artificial intelligence (AI) have raised concerns about their potential to create convincing fake social media accounts, but empirical evidence is lacking.
In this paper, we present a systematic analysis of Twitter (X) accounts using human faces generated by Generative Adversarial Networks (GANs) for their profile pictures. 
We present a dataset of 1,420 such accounts and show that they are used to spread scams, spam, and amplify coordinated messages, among other inauthentic activities.
Leveraging a feature of GAN-generated faces---consistent eye placement---and supplementing it with human annotation, we devise an effective method for identifying GAN-generated profiles in the wild.
Applying this method to a random sample of active Twitter users, we estimate a lower bound for the prevalence of profiles using GAN-generated faces between 0.021\% and 0.044\%---around 10K daily active accounts.
These findings underscore the emerging threats posed by multimodal generative AI.
We release the source code of our detection method and the data we collect to facilitate further investigation.
Additionally, we provide practical heuristics to assist social media users in recognizing such accounts.
\end{abstract}

\section{Introduction}

Recent advances in generative artificial intelligence (AI) have resulted in remarkable competencies across various domains~\citep{cao2023comprehensive}.
Large language models, such as ChatGPT,\footnote{\url{https://chat.openai.com}} can understand human language and generate realistic text in different contexts~\citep{jakesch2023human,ye2023comprehensive}.
Generative vision models, such as Generative Adversarial Networks (GANs)~\citep{goodfellow2014generative} and diffusion models~\citep{song2019generative,ho2020denoising}, produce photorealistic images that people can no longer distinguish from human-generated images~\citep{lago2021more,nightingale2022ai,bray2023testing}.
State-of-the-art multimodal models, such as Google's Gemini models~\citep{gemini2023}, can even understand visual and audio inputs and respond with text or images.
This evolving landscape of generative AI models underscores a significant leap in AI technology, opening new avenues for research and applications.

These advancements also present opportunities for misuse.
For instance, adversarial actors can exploit AI models to create and disseminate disinformation or conduct influence campaigns~\citep{yamin2021weaponized,guembe2022emerging,goldstein2023generative,yang2024anatomy}, threatening the integrity of online information exchange.
In this study, we focus on the fake social media accounts that leverage generative AI for deceptive purposes~\citep{bommasani2021opportunities}. 
In documented cases, individuals have used AI-generated human images to fabricate fake personas online, leading to deception and targeted attacks~\citep{osullivan2020fake,osullivan2020high}.
Major social media platforms, such as Facebook~\citep{martineau2019facebook} and LinkedIn~\citep{bond2022smiling}, have already encountered groups of fake accounts that utilize AI-generated faces. 

Despite these stories, a systematic analysis of such accounts is still lacking.
What are the tactics and goals of these fake accounts?
What is their prevalence on social media platforms?
These key questions remain unanswered due to the reluctance of social media platforms to share data and detailed analysis results.
To address this knowledge gap, here we present an analysis of inauthentic accounts using GAN-generated pictures as their profiles to mimic real human users on Twitter.\footnote{Twitter was rebranded as ``X'' in July 2023, but we refer to the platform as ``Twitter'' as the data analyzed in this paper predates the rebranding.}

Although there are other generative vision models, we focus on GANs for several reasons.
Firstly, the introduction of GAN-based models, particularly the StyleGAN series~\citep{karras2019style,karras2020analyzing}, marks a major milestone in the field of generative AI models~\citep{cao2023comprehensive} because they are capable of producing photorealistic images almost indistinguishable from real ones~\citep{nightingale2022ai}.
Secondly, GANs are easily accessible.
With their source code publicly available online,\footnote{See \url{https://github.com/NVlabs/stylegan} and \url{https://github.com/NVlabs/stylegan2}.} adversarial actors can easily produce fake human faces as needed.
Additionally, for those without technical expertise, websites like \url{https://thispersondoesnotexist.com} offer an easy means to download GAN-generated images in large quantities.
Last but not least, GAN-generated faces share some distinctive features, allowing us to identify them with high confidence.

In this study, we create a novel dataset of 1,420 inauthentic Twitter accounts, each utilizing a GAN-generated face as its profile picture.
Upon close examination, these accounts typically feature human-like names suggesting a deliberate effort to mimic real users.
These accounts engage in inauthentic activities, such as disseminating spam, executing scams, and amplifying messages in a coordinated fashion, often through automation.
We further estimate the prevalence of such accounts on the platform.
Leveraging the feature that GANs consistently place eyes at the same locations, together with human annotation, we develop a simple yet effective method to identify GAN-generated profiles in the wild.
Applying this method to a random sample of 254,275 active Twitter users, our analysis indicates that the prevalence of accounts using GAN-generated faces is between 0.021\% and 0.044\%. 
Since our method emphasizes precision and might miss accounts leveraging more advanced AI models to generate their profiles, these estimates should be only regarded as lower bounds.

Our findings reveal that generative AI tools, particularly GANs, have been deployed at scale to create fake personas on Twitter.
These accounts are involved in various inauthentic activities, posing a challenge to the integrity of online interactions.
To support further research and expand the scope of investigation to other cases and platforms, we share our dataset and code.
We also discuss practical heuristics that can empower social media users to identify and defend themselves against the manipulative tactics of these AI-powered accounts.

\section{Related work}

\subsection{GAN-generated face detection}

Due to the popularity of the StyleGAN models~\citep{karras2019style,karras2020analyzing} and their broad impact, many studies have aimed to detect GAN-generated pictures in the past few years.
In addition to deep learning models that can capture distinctive features of GAN-generated pictures automatically~\citep{nataraj2019detecting,dang2018deep}, many methods rely on physical or physiological features of the generated faces for detection.
For example, \citet{hu2021exposing} leverage corneal specular highlights; \citet{guo2022eyes} focus on pupil shapes; \citet{yang2019exposing} and \citet{mundra2023exposing} exploit the locations of face landmarks.
For details on these methods, readers can refer to the review by \citet{wang2022gan}.

Although we also need to detect GAN-generated faces in this study, our case differs from previous studies significantly.
The aforementioned methodology papers typically aim to distinguish generated and real human faces from datasets such as FFHQ~\citep{karras2019style} and CelebA~\citep{liu2015deep}. 
On the other hand, our objective is to identify generated faces among random Twitter profiles, which presents unique challenges.
Firstly, only a subset of Twitter profiles feature human faces, and their styles may differ greatly from GAN-generated ones (see Figure~\ref{fig:samples_tiled}(c) for examples).
Secondly, the instances of GAN-generated faces may be relatively scarce, necessitating a detection method with high accuracy to address the class imbalance issue and provide a reliable estimate of the prevalence of fake accounts.
Given these differences, existing methods are not suitable for our purpose.

\subsection{GAN-generated faces on social media}

While there is a substantial body of literature on detecting GAN-generated images, only a few studies have specifically targeted social media contexts.
\citet{marra2018detection} simulate the case of detecting generated pictures compressed by social media platforms by preprocessing the pictures in their datasets, although their focus is not on human faces.
\citet{boato2022trueface} upload GAN-generated human pictures to social media platforms, including Facebook, Twitter, and Telegram, and then download them to examine the effectiveness of various detection methods on compressed images.

Unfortunately, these studies cannot help address our research questions about the fake social media accounts that utilize GAN-generated faces.
Our knowledge of these accounts is largely derived from sporadic journalistic reporting, brief communications from social media platforms, and case studies provided by hobbyists~\citep{norteno2023history}.
We hope the present study can further our understanding of these accounts and their negative impact.

\subsection{Social bots and their detection}

This study also encompasses the topic of social bots, which, while related, are distinct from our primary focus. 
Social bots are accounts partially controlled by algorithms to emulate humans and act automatically following different instructions~\citep{ferrara2016rise}.
To appear authentic, bots sometimes employ realistic images as profile pictures, including those stolen from other accounts~\citep{confessore2018follower} or generated using AI models (as we show below).
However, accounts using GAN-generated faces are not necessarily bots; sometimes, they are operated by real humans.

The detection of social bots often depends on the content produced by the accounts and their behavioral patterns, such as posting frequency and types of interactions with other accounts~\citep{yang2019arming,botometerv4-2020}.
Some recent studies have explored using GANs for bot detection in adversarial settings~\citep{cresci2021coming,najari2022ganbot}.
However, their focus is mainly on the behavioral patterns and content, not on the profile pictures.
Account profile information, such as the user name, description, and number of friends and followers, typically contributes to bot detection~\citep{yang2020scalable}, but profile pictures are usually excluded, likely due to image-processing challenges.

Since we aim to identify accounts that use GAN-generated faces in this study, we will mainly rely on profile pictures for detection.
We will then analyze the profile information and activities of the identified accounts.

\section{Data collection}

We introduce multiple datasets for different purposes.
The key dataset is shared with the public for further investigation and can be accessed at a public data repository.\footnote{\url{https://zenodo.org/doi/10.5281/zenodo.10436888}}

\subsection{TwitterGAN dataset}

\begin{table*}
    \centering
    \caption{Detailed information about different groups of accounts in \dataset{TwitterGAN}.}
    \resizebox{\textwidth}{!}{
    \begin{tabular}{lrlp{8cm}}
    \toprule
    Group name & Size & Date & Note\\
    \midrule
    blue\_check & 11 & 2022-11-10  & Subscribed to the blue-checkmark paid verification program of Twitter \\
    \addlinespace
    bart& 5 & 2023-02-02 & Used to propagate spammy hashtags, YouTube videos, and Tumblr links\\
    \addlinespace
    elon\_watch & 246&  2023-02-03 & Repeatedly replied to other accounts with scams involving Elon Musk \\
    \addlinespace
    sunday & 18 &  2023-03-05 & Used to amplify the tweets of Twitter user @ilkersenock\\
    \addlinespace
    astroturf & 1,066 &  2023-04-08 & Used to amplify content from a common set of accounts in a coordinated fashion\\
    \addlinespace
    chatgpt & 67 & 2024-01-31 &  Used ChatGPT to generate human-like content\\
    \addlinespace
    ind\_acc & 7 &  varies  & Individual accounts that do not belong to any coordinated groups\\
    \midrule
    Total & 1,420 & &  \\
    \bottomrule
    \end{tabular}
    }
    \label{tab:twitter_gan}
\end{table*}

The \dataset{TwitterGAN} dataset comprises a total of 1,420 Twitter accounts that use GAN-generated human faces as their profiles.
We collected their profile information and recent tweets for analysis.
When Twitter's V2 API was still available, we used it to collect the data for most of the accounts.
After Twitter shut down its free API, we manually collected the data through Twitter's web interface.
We also downloaded their profile pictures.
Twitter allowed public access to these pictures in various resolutions.
We primarily aimed for the 400-by-400-pixel versions, although some variations in resolution and aspect ratios exist.

The dataset contains multiple account groups collected between November 2022 and January 2024.
The timeline of data collection and notes can be found in Table~\ref{tab:twitter_gan}.
The initial identification of some of these accounts was done by the Twitter user @conspirator0, while others were spotted by the authors.
Following this, we conducted a thorough inspection to verify the nature of these accounts.

\begin{figure}
    \centering
    \includegraphics[width=\textwidth]{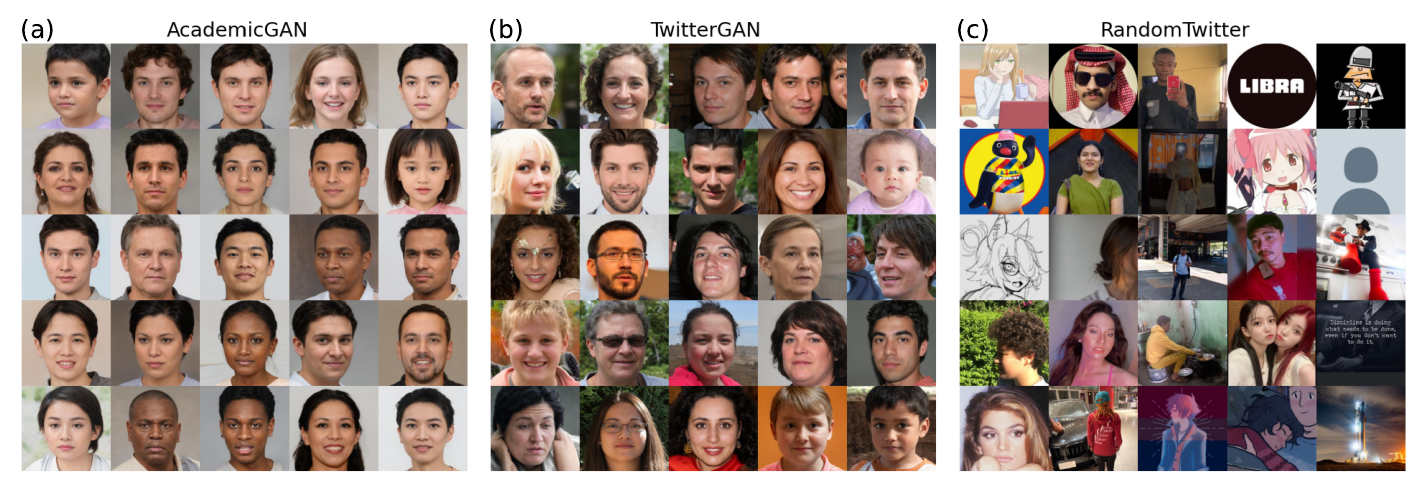}
    \caption{
    Sample profile pictures from (a) \dataset{AcademicGAN}, (b) \dataset{TwitterGAN}, and (c) \dataset{RandomTwitter}.
    }
    \label{fig:samples_tiled}
\end{figure}

As mentioned earlier, the GAN-generated faces are very realistic, and layman participants in controlled experiments can no longer distinguish them from real human faces~\citep{nightingale2022ai}.
However, after reading extensive literature and reviewing a large number of GAN-generated human faces, we have identified several distinctive characteristics of these faces that bolster our confidence in their identification.
First, GANs consistently show face landmarks, especially the eyes, at the same location in the generated pictures~\citep{yang2019exposing}.
This feature becomes evident when comparing multiple GAN-generated faces side by side, as illustrated in Figure~\ref{fig:samples_tiled}.
Second, many GAN-generated pictures have clear defects, which will be discussed in Section~\ref{sec:heuristics}.
Third, some profile pictures contain watermarks explicitly revealing their machine-generated origin.

We also examine the profile information and activities of the accounts to triangulate their nature.
For instance, there are occasional mismatches between account names and the perceived gender and race of the faces.
We further rely on signals of coordinated inauthentic activities, which proved effective in similar contexts~\citep{pacheco2020uncovering,yang2024anatomy}. 
The \dataset{TwitterGAN} accounts in some groups are inter-connected through mutual following, similar profile information, and identical activity patterns. (These connections also allow us to identify more members of their groups.) 
Our conclusions are drawn from a holistic consideration of all these factors.

\subsection{AcademicGAN dataset}

Since capturing a large number of accounts using GAN-generated profiles in the wild is not trivial, we supplement the study with the \dataset{AcademicGAN} dataset.
It contains 10,000 GAN-generated human faces obtained from the \url{https://generated.photos/datasets} website upon request.
The website trains its own GAN models on proprietary human photos to produce more synthetic ones for sale.
All pictures in \dataset{AcademicGAN} have the same resolution of 256 by 256 pixels and do not have the common defects observed in \dataset{TwitterGAN}.
Nevertheless, their face landmarks are still placed at the same locations, as demonstrated in Figure~\ref{fig:samples_tiled}.
This dataset is used as the ground truth in our analysis to compare with \dataset{TwitterGAN} and develop our detection method.

\subsection{RandomTwitter dataset}

To estimate the prevalence of accounts using GAN-generated profiles on Twitter, we assemble the \dataset{RandomTwitter} dataset, a random sample of diverse active Twitter users.
We leverage the streaming endpoint of Twitter's V2 API, which delivers a 1\% sample of all live tweets in real time.
The data collection took place on June 1, 2023, before Twitter shut down our API access. 
This yielded 285,747 tweets from 254,275 unique users.
Similar to the \dataset{TwitterGAN} dataset, we also downloaded the 400-by-400-pixel versions of their profile images whenever possible.
Figure~\ref{fig:samples_tiled} shows a sample of 25 of these images.

\section{Characterization of TwitterGAN}

In this section, we characterize the profiles and activities of accounts in \dataset{TwitterGAN}.

\subsection{Profile information}

\begin{figure}
    \centering
    \includegraphics[width=0.8\columnwidth]{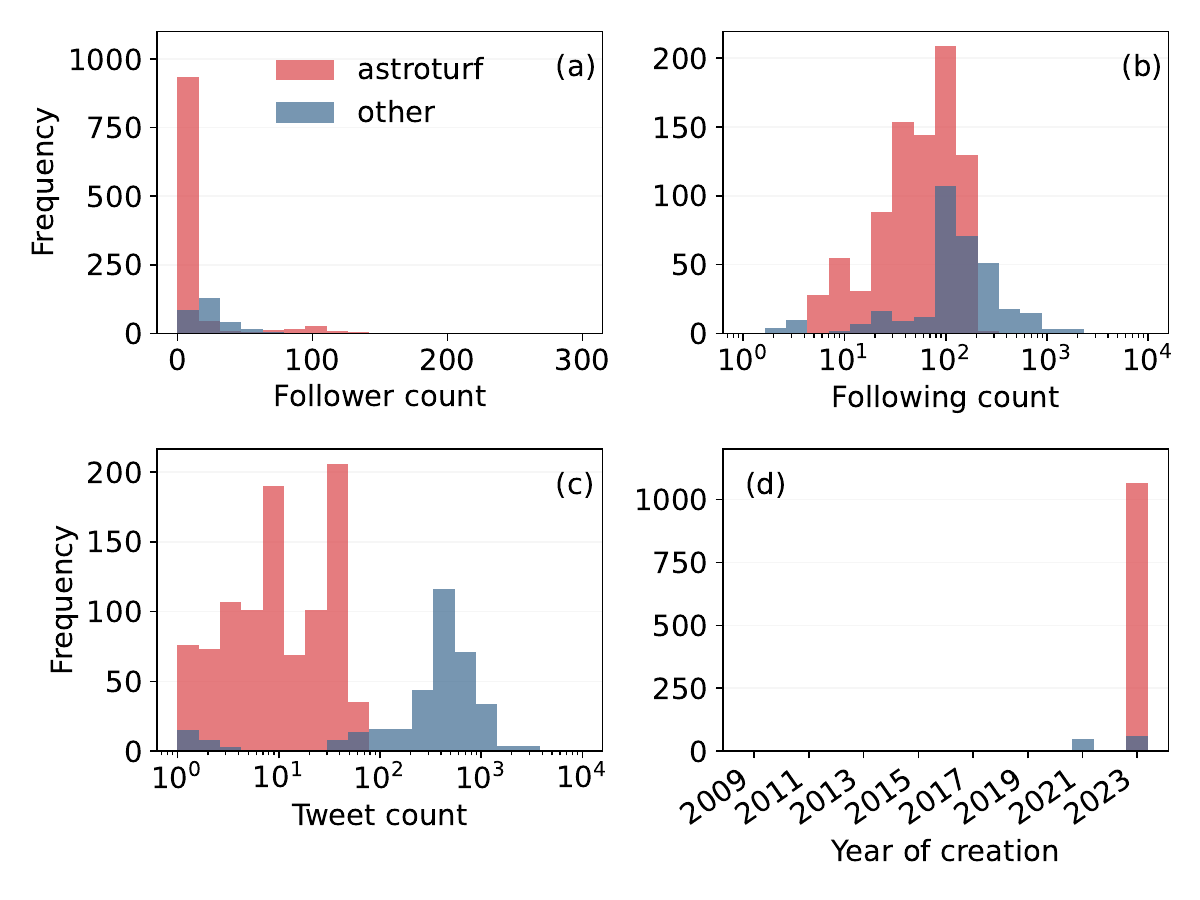}
    \caption{
    Profile characteristics of accounts in \dataset{TwitterGAN}.
    We show the distributions of (a) follower count, (b) following (friend) count, (c) tweet count, and (e) year of creation for \dataset{astroturf} and other accounts.
    }
    \label{fig:account_char}
\end{figure}

Let us first focus on the profile information of the accounts in \dataset{TwitterGAN}.
The distributions of their follower/following count, tweet count, and creation year are shown in Figure~\ref{fig:account_char}.
Since the \dataset{astroturf} group has more accounts than the other groups combined and shows distinctive features, we separate it from the rest of the accounts.

On average, accounts in \dataset{astroturf} have 9.95 (Standard Deviation SD=24.03) followers, 55.68 (SD=51.75) friends, and 14.91 (SD=15.62) tweets, whereas the other accounts have 67.75 (SD=163.64) followers, 162.71 (SD=177.17) friends, and 509.23 (SD=504.02) tweets.
All the differences between \dataset{astroturf} and the other accounts are statistically significant according to Kolmogorov–Smirnov tests ($p<0.05$).
These results suggest that accounts in \dataset{TwitterGAN} are engaging in various activities on Twitter; however, \dataset{astroturf} accounts are less active than others.

Inspecting the creation years of the accounts suggests that most of them joined Twitter in the past two years, and \dataset{astroturf} accounts are more recent than others.
This is not surprising since Twitter, both under the old management and Musk, routinely removes inauthentic accounts identified by their algorithm or reported to them.
But this process is imperfect since some accounts in \dataset{TwitterGAN} are still alive on Twitter despite violating the policies.
Specifically, all accounts in \dataset{astroturf} appeared to be suspended when we checked their status in December 2023, but only 14.9\% of the other accounts in \dataset{TwitterGAN} had been suspended.

\subsection{Tactics and inauthentic activities}

\begin{figure}[t]
    \centering
    \includegraphics[width=0.8\textwidth]{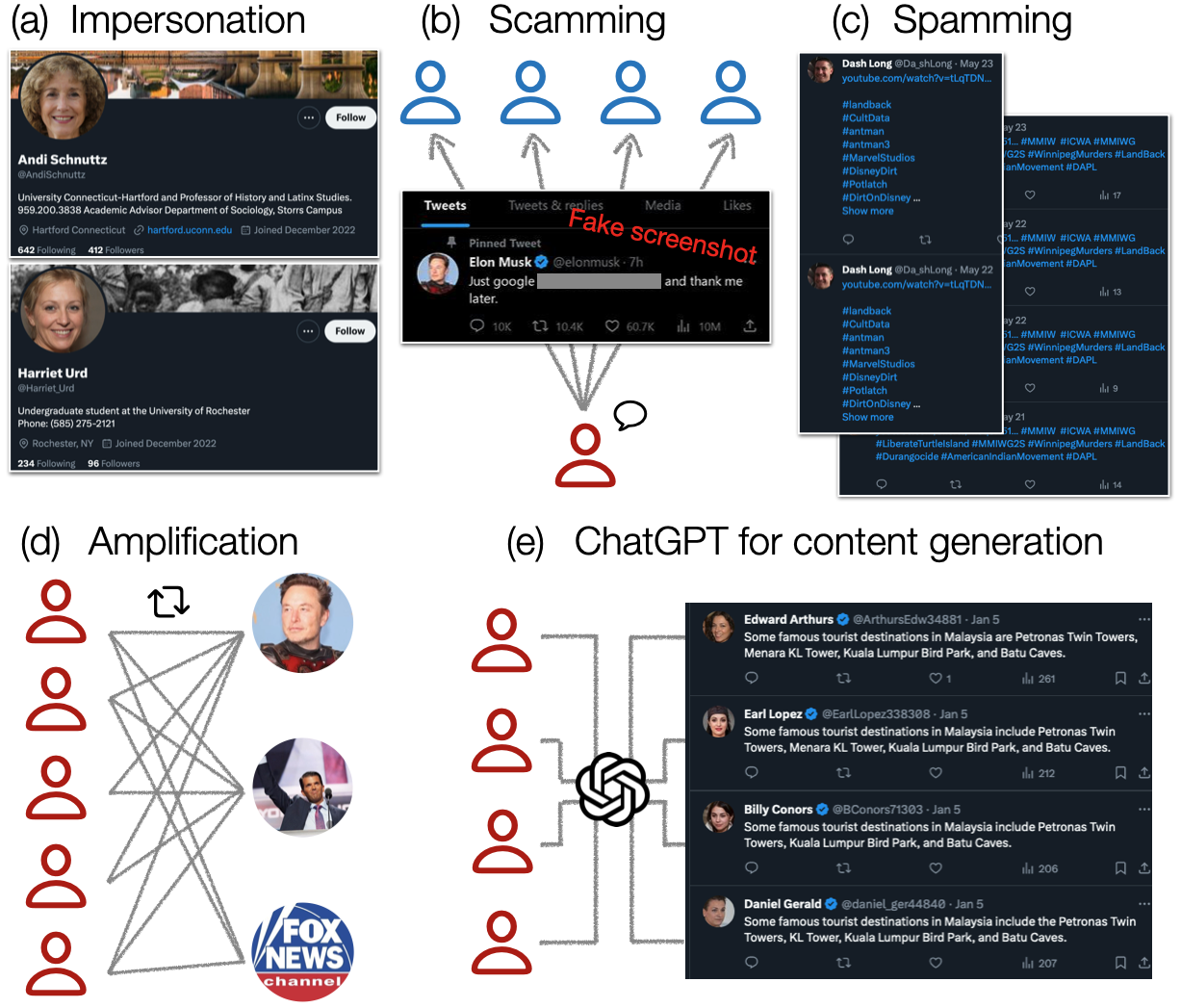}
    \caption{
    Illustrations of the tactics and inauthentic activities of accounts in \dataset{TwitterGAN}.
    (a) Fake accounts impersonating persons that do not exist through GAN-generated profiles, human-like names, and fake descriptions.
    (b) Accounts replying fake screenshots with scam messages to other accounts (the keywords are redacted to avoid further spreading the scams).
    (c) Accounts posting spam messages repeatedly.
    (d) Accounts amplifying the messages from certain accounts by coordinately reposting their tweets.
    (e) Accounts leveraging ChatGPT to produce human-like replies.
    The account illustrations colored in red and blue represent inauthentic and normal ones, respectively.
    }
    \label{fig:tactics}
\end{figure}

To uncover the tactics and inauthentic activities of accounts in \dataset{TwitterGAN}, we performed human annotation and bot detection.
The annotation covered all accounts from \dataset{blue\_check}, \dataset{bart}, \dataset{elon\_watch}, \dataset{sunday}, \dataset{chatgpt}, and \dataset{ind\_acc}.
Since the accounts in \dataset{astroturf} share similar patterns, we randomly selected 250 for annotation to reduce the workload.
In total, we annotated 604 accounts, and each account was annotated by two authors independently.
Our analyses reveal several patterns, and we illustrate the five most noteworthy ones in Figure~\ref{fig:tactics}.
Some patterns are common across multiple groups, while others are exclusive to specific ones. 
We discuss the patterns next.

\begin{description}

\item[Impersonation.]

The primary goal of using GAN-generated images in the profiles is to impersonate real humans.
As illustrated in Figure~\ref{fig:tactics}(a), these fake accounts often feature realistic photos, plausible names, and detailed descriptions, creating the illusion that these accounts are operated by real persons.
Our analysis suggests that almost all \dataset{TwitterGAN} accounts employ believable names despite some being uncommon. 
None of these accounts disclose their use of GAN-generated profile images.

\item[Scamming.]

Our analysis suggests that the accounts in the \dataset{elon\_watch} group are used to promote scams.
These accounts typically reply to others with screenshots depicting Elon Musk's Twitter feed, containing deceptive messages like ``Just google [keywords] and thank me later.''
An illustration can be found in Figure~\ref{fig:tactics}(b).
The screenshots are fabricated; Elon Musk never posted these messages.
Searching the keywords on Google leads to scam websites. 

\item[Spamming.]

Accounts within \dataset{TwitterGAN} frequently engage in spamming.
We show two such examples from the \dataset{bart} group in Figure~\ref{fig:tactics}(c), characterized by repeating messages with excessive hashtags and external links to platforms like YouTube and Tumblr.

\item[Coordinated amplification.]

Some of the \dataset{TwitterGAN} accounts exhibit coordinated inauthentic activity (see Figure~\ref{fig:tactics}(d)), a tactic commonly adopted by adversarial actors during influence campaigns~\citep{pacheco2020uncovering}.
A clear example is the \dataset{astroturf} group, where the members retweet content from a common set of accounts.
The five most retweeted accounts are: \account{elonmusk} (Elon Musk's account), \account{DonaldJTrumpJr} (Donald Trump Jr.'s account), \account{FoxNews} (the official account of Fox News), \account{POTUS} (the official account of the US President, currently owned by Joe Biden), and \account{Crunchyroll} (the official account of \url{https://crunchyroll.com}, an anime distribution company).
Their motives remain unclear, given the varied nature of the amplified content.

\item[ChatGPT for content generation.]

Accounts in \dataset{chatgpt} were found to leverage ChatGPT to generate human-like content (see Figure~\ref{fig:tactics}(e)).
They were identified by the self-revealing tweets posted by these accounts, which contained phrases such as ``as an AI language model.''
This method was used to uncover a larger botnet with the same tactic earlier~\citep{yang2024anatomy}.

\item[Automation.]

Many accounts in \dataset{TwitterGAN} exhibit characteristics of bots.
For instance, repeatedly sending scams, posting spam messages, and amplifying posts from selected accounts in a coordinated fashion are typical signs of automation.
We further apply the BotometerLite model~\citep{yang2020scalable}, a widely adopted Twitter bot detection tool, to all accounts in \dataset{TwitterGAN} and find that 75.2\% of them are labeled likely bots (using 0.5 as threshold).

\item[``Verification.'']

Under Musk's management, the acquisition of a blue checkmark no longer requires going through a verification process.
Instead, any account willing to pay a premium fee can obtain the mark to increase its perceived legitimacy. 
This strategy is not very common among \dataset{TwitterGAN} accounts, possibly due to the associated cost.
Only the 11 accounts in \dataset{blue\_check} and 44 accounts in \dataset{chatgpt} adopted this strategy.

\end{description}

\subsection{Heuristics for identification}
\label{sec:heuristics}

\begin{figure}
    \centering
    \includegraphics[width=0.5\columnwidth]{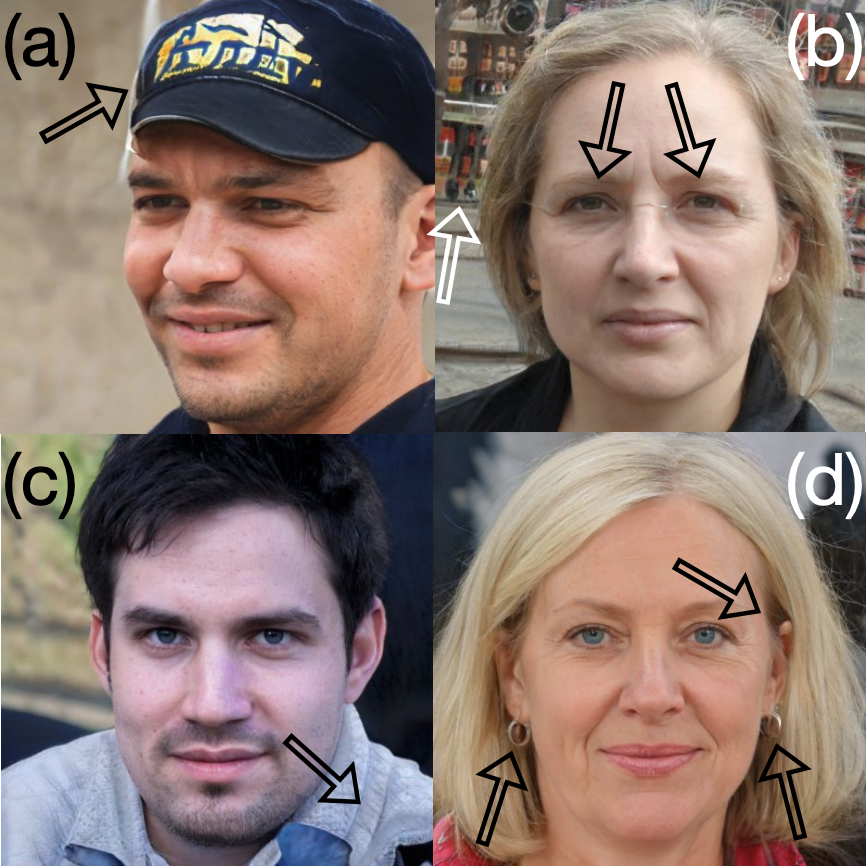}
    \caption{
    Common distinctive defects of GAN-generated faces that can help identify them. 
    (a) Unreal hat. (b) Glass frame blended into the face; surreal background. (c) Unreal collar. (d) Earrings blended into the ears; irregular ear.
    }
    \label{fig:heuristics}
\end{figure}

During the annotation, we identified some heuristic indicators that can help spot GAN-generated profiles.
The most distinctive feature lies in the flaws.
Although GANs can generate realistic human faces, other parts of the images are often featured with defects.
For instance, GANs sometimes have difficulty producing realistic hats (Figure~\ref{fig:heuristics}(a)), glasses (Figure~\ref{fig:heuristics}(b)), and ears (Figure~\ref{fig:heuristics}(d)).
Similarly, users can pay attention to accessories, such as earrings (Figure~\ref{fig:heuristics}(d)) and necklaces, that GANs often struggle with.
Clothing details, especially collars, are also problematic for GANs (Figure~\ref{fig:heuristics}(c)).
Last but not least, the background sometimes can hint at the artificial nature of a picture (Figure~\ref{fig:heuristics}(b)).
While annotating \dataset{TwitterGAN} accounts, we specifically examined the defects in their profile pictures.
Two annotators agreed that 60.1\% of them have clear defects (Cohen's Kappa = 0.90).

% Inconsistency
Profile images lacking obvious defects can still be identified through inconsistencies between images and profile information.
For instance, there may be a disconnect between the perceived gender or ethnicity in the image and the associated names. 
Such mismatches often result from creators randomly pairing images and names without validation.
However, relying solely on this criterion can be inaccurate and potentially biased, so it should be used cautiously and complemented with additional verification.

% Coordinated behaviors
Although it is challenging to determine the nature of suspicious accounts individually, patterns emerge when comparing multiple accounts. 
Users can start with one suspicious account and observe its connections and interactions with others to search for signs of coordinated inauthentic behaviors.
These include accounts sharing compatible profile pictures and/or exhibiting similar activities, such as posting similar content and retweeting similar posts.
Such analysis can contextualize individual accounts within larger networks, aiding in identifying their true nature.

\section{Prevalence on Twitter}

In this section, we aim to estimate the prevalence of Twitter accounts using GAN-generated profiles.
We first propose a simple yet effective detection method, then apply it to \dataset{RandomTwitter}.

\subsection{Eye locations}

Our detection method leverages the observation that GANs place the eyes in the same locations in all generated pictures, as demonstrated in Figure~\ref{fig:samples_tiled}(a,b).
This feature has been reported and exploited to distinguish true and fake human faces~\citep{yang2019exposing}.
Note that our task differs from previous methodology papers since our goal is to identify GAN-generated profiles from random Twitter profiles.
As illustrated in Figure~\ref{fig:samples_tiled}(c), many profiles do not feature human faces.
And for those containing authentic human faces, we hypothesize that very few of them happen to have their eyes placed in the same locations as the GAN-generated ones.

To test our hypothesis, we need to obtain the eye coordinates of input images.
We test different face recognition software and find that the Python package \package{face\_recognition}\footnote{\url{https://github.com/ageitgey/face_recognition}} provides an effective and efficient solution.
The package returns an empty list if no faces are detected in the input images.
For images containing faces, the package returns the coordinates of the landmarks, such as the eyes and nose, of each face~\citep{kazemi2014one}.
We focus on the eyes and obtain the center coordinate of each eye by averaging its landmark points.
Since the profile images have different resolutions, we normalize the coordinates using the image height and width so that all the re-scaled coordinates are in the unit interval.

\begin{figure}[t]
    \centering
    \includegraphics[width=\columnwidth]{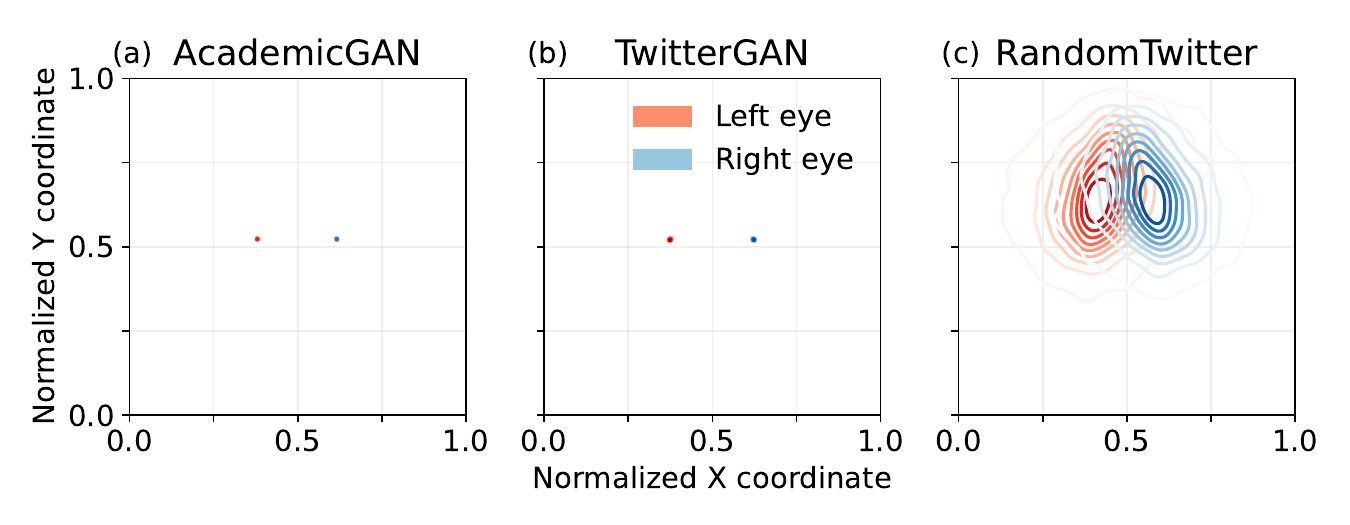}
    \caption{
    Kernel density estimation distribution of the detected left and right eyes in (a)~\dataset{AcademicGAN}, (b)~\dataset{TwitterGAN}, and (c)~a random sample of 25,000 accounts from \dataset{RandomTwitter}.
    }
    \label{fig:eye_location_dist}
\end{figure}

We apply the package to the profiles in \dataset{TwitterGAN}, \dataset{AcademicGAN}, and \dataset{RandomTwitter}.
Figure~\ref{fig:eye_location_dist} shows the kernel density estimation distributions of the detected eye positions.
For the GAN-generated profiles in both \dataset{TwitterGAN} and \dataset{AcademicGAN}, the eye locations are highly concentrated and identical.
On the other hand, the eye locations of random profiles in \dataset{RandomTwitter} are scattered across the space.
These findings confirm our hypothesis and suggest that  TwitterGAN images were likely generated by the same algorithm as those in AcademicGAN. In turn, this indicates eye locations can be leveraged to distinguish GAN-generated and random Twitter profiles. 

Inspecting the results of \package{face\_recognition} yields a few more critical observations.
The package can identify exactly one face from each \dataset{AcademicGAN} image and each \dataset{TwitterGAN} image, except for one with clear defects.
On the other hand, the package can only identify faces from around 35.7\% profiles in \dataset{RandomTwitter}.
The remaining profiles either do not contain faces or are misclassified by the package.
Among the profiles with detected faces, about 92.3\% have one face detected, while the others have multiple faces detected.
Although \package{face\_recognition} is not perfect in face detection, it works very well on the GAN-generated faces, ensuring a high recall for our objective.

\subsection{GANEyeDistance metric}

Based on the promising results in Figure~\ref{fig:eye_location_dist}, we propose a metric called GANEyeDistance to quantify the distances between the detected eyes and the expected coordinates of the eyes in GAN-generated pictures.

Let us use $\vec{L} = \{x_l, y_l\}$ and $\vec{R} = \{x_r, y_r\}$ to represent the coordinates of the left and right eyes detected by \package{face\_recognition}, where $x$ and $y$ represent the normalized x- and y-coordinates.
We further use $\vec{L}_{\text{GAN}}$ and $\vec{R}_{\text{GAN}}$ to denote the expected locations of left and right eyes in GAN-generated images.
The GANEyeDistance for a given profile is defined as:
\begin{equation}
\mathcal{G} (\vec{L}, \vec{R}) =
    \begin{cases}
        \frac{\lVert \vec{L} - \vec{L}_{\text{GAN}} \rVert + \lVert \vec{R} - \vec{R}_{\text{GAN}} \rVert}{2 \sqrt{2}} & \text{if one detected face}\\
        1 & \text{otherwise}
    \end{cases} \label{eq:ganed}
\end{equation}
where $\lVert \vec{X} - \vec{Y} \rVert$ represents the Euclidean distance between $\vec{X}$ and $\vec{Y}$.
Since the longest distance of two points in a unit square is $\sqrt{2}$ and we combine the distance of both eyes, the denominator in Eq.~(\ref{eq:ganed}) ensures that $\mathcal{G}$ is a number between 0 and 1.
A profile with a $\mathcal{G}$ value close to 0 means the eye locations are close to those of GAN-generated faces.

Note that $\mathcal{G}$ is only calculated when one face is detected in the given profile.
If \package{face\_recognition} cannot process the input image or cannot detect any faces, we set $\mathcal{G}$ to 1.
We also set $\mathcal{G}$ to 1 when multiple faces are detected since GAN-generated profiles typically have one face in them.

To apply Eq.~(\ref{eq:ganed}), we also need the ground truth values of $\vec{L}_{\text{GAN}}$ and $\vec{R}_{\text{GAN}}$.
Inspecting the profiles in \dataset{TwitterGAN} suggests that the profile pictures are subject to modifications such as cropping and stretching, introducing noise.
On the other hand, pictures in \dataset{AcademicGAN} have better consistency and quality, so we use the average values of their eye coordinates as our ground truth.
Our implementation of GANEyeDistance is publicly available.\footnote{\url{https://github.com/osome-iu/fake_gan_accounts}}

\begin{figure}[t]
    \centering
    \includegraphics[width=\columnwidth]{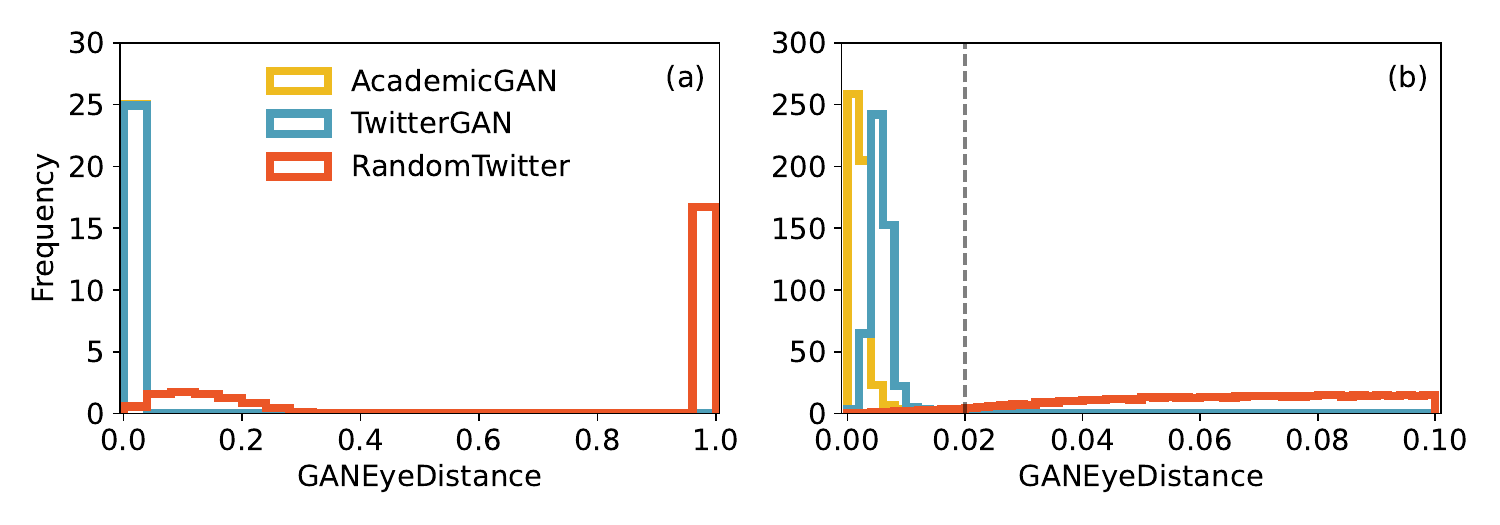}
    \caption{
    (a) Distributions of the GANEyeDistance values in \dataset{AcademicGAN}, \dataset{TwitterGAN}, and \dataset{RandomTwitter}.
    The results for \dataset{AcademicGAN} and \dataset{TwitterGAN} overlap with each other.
    (b) A zoom-in of (a) with a focus on the range of $[0, 0.1]$.
    The dashed line indicates the threshold we choose for identifying likely GAN-generated profiles in \dataset{RandomTwitter}.
    }
    \label{fig:ganed_dist}
\end{figure}

To demonstrate the effectiveness of GANEyeDistance, we calculate this value for every image in all three datasets and show their distributions in Figure~\ref{fig:ganed_dist}(a).
We can see that the profiles in \dataset{AcademicGAN} and \dataset{TwitterGAN} have GANEyeDistance values very close to 0, as expected.
The results of \dataset{RandomTwitter} show a bimodal distribution: the GANEyeDistance values of the profiles with one detected face are distributed across a wide range, and the remaining images have a GANEyeDistance value of 1.
The difference in the distributions suggests that GANEyeDistance can effectively distinguish GAN-generated profiles from random Twitter profiles.
In Figure~\ref{fig:ganed_dist}(b), we zoom in to better inspect the difference among the three datasets.
The GANEyeDistance values for \dataset{AcademicGAN} are closest to 0.
The results for \dataset{TwitterGAN} deviate from 0 slightly due to noise.
The \dataset{AcademicGAN} distribution has a very small overlap with those of the GAN-generated profiles.

Despite the promising results shown in Figure~\ref{fig:ganed_dist}, we emphasize again that GANEyeDistance only indicates the distance between the locations of detected eyes and the ground truth.
It cannot truly distinguish GAN-generated faces from authentic human faces.
This means that although GAN-generated faces have GANEyeDistance close to zero, images with very small GANEyeDistance are not necessarily GAN-generated.
In other words, a relatively loose threshold for GANEyeDistance can yield a very high recall for GAN-generated profiles but also introduce false positives.

For instance, if we use 0.02 as the threshold and consider all images with $\mathcal{G}<0.02$ as positive cases, then the recall for \dataset{AcademicGAN} and \dataset{TwitterGAN} is 100\% and 99.4\%, respectively.
At the same time, 0.46\% (1,181) of the random profiles are labeled as positive, so further inspection is needed to determine their nature.

\subsection{Prevalence estimation}

In this study, our goal is not to build a perfect algorithm to detect GAN-generated profiles.
Instead, we aim to estimate the prevalence of accounts using GAN-generated profiles on Twitter, which can be approximated by the prevalence in \dataset{RandomTwitter}.
Therefore, the high recall nature of GANEyeDistance makes it a suitable metric to help us narrow down the candidate profiles for further inspection.

Here, we settle on 0.02 as the threshold because it yields a very high recall for \dataset{TwitterGAN}.
Increasing the threshold does not boost the recall significantly but introduces many more false positives.
As mentioned above, this threshold leads to 1,181 candidate profiles from \dataset{RandomTwitter}.

To reduce the false positives where the algorithm flags authentic human photos as GAN-generated, two authors manually classify these profiles into three categories: (1) highly likely GAN-generated, (2) likely GAN-generated, and (3) not GAN-generated.
Profiles in the first category demonstrate clear and unique features of GAN-generated human faces, such as the defects illustrated in Figure~\ref{fig:heuristics}.
Profiles in the second category are very similar to GAN-generated faces; however, no definitive evidence can be identified.

The Cohen's Kappa for the ratings is 0.85, indicating high agreement between the two annotators.
54 profiles are coded as highly likely generated by GANs by both annotators, and 113 are believed to be likely or highly likely generated by GANs by both annotators.
Profiles are considered not GAN-generated if at least one annotator believes so.
Considering that these profiles are selected from the 254,275 accounts in \dataset{RandomTwitter} and our method might miss some positive cases, we estimate a lower bound on the prevalence of active Twitter users using GAN-generated profiles to be between 0.021--0.044\%.
These accounts were responsible for 0.022--0.053\% of the tweets in \dataset{RandomTwitter}.

\section{Discussion}

In summary, we present a systematic analysis of fake accounts using GAN-generated faces on Twitter.
We show that they impersonate humans with realistic profiles and are involved in various inauthentic activities, ranging from spamming and scamming to coordinated amplification of messages.
We also provide some heuristics that social media users can use to identify the accounts they encounter.

Using a random sample, we further estimate the prevalence of such accounts on Twitter.
We first build a heuristic method exploiting the common feature of GAN-generated faces to narrow down the potential GAN-generated profiles, then perform manual annotation to identify those likely to be synthetic.
Our results suggest that the percentage of active Twitter accounts using GAN-generated profiles is between 0.021\% and 0.044\%.
To put the percentages into perspective, let us consider the total number of daily active Twitter users.
\citet{pfeffer2023just} collected all 375 million tweets posted on September 21, 2022, and found 40,199,195 active users.
Using this number, we estimate that at least 8,537--17,864 daily active accounts on Twitter use GAN-generated profiles.

Since the recall of our machine learning model is not 100\%, and GAN-generated images that are heavily stretched or tilted can evade our method, we might miss some positive cases in our analysis.
Also, the labeling criteria used in our annotation process are very conservative---we only consider a profile positive when two annotators are both highly certain about the label.
Therefore, these estimates should be considered as a lower bound. 
In fact, a recent report by \citet{ricker2024aigenerated} produced a similar but slightly higher estimate of 0.052\% using different detection methods and excluding accounts with default profile images in the denominator.

Our analysis and estimation have some limitations.
We only include a few groups of accounts in \dataset{TwitterGAN}, so their behavior patterns might not generalize to other cases.
Although we have proposed an effective method to capture more accounts using GAN-generated profiles in the wild, further analysis is not feasible since Twitter ended free access to its API.
Also, since we only focus on one platform, the findings might not generalize to other platforms, such as Facebook and LinkedIn, that have been reported to suffer from the same issue. 
Unfortunately, limited access to their data makes it impossible for researchers to study inauthentic accounts on those platforms.
The Digital Services Act by the European Union mandates that large social media platforms, including Twitter and Meta, provide data access to researchers, bringing hope.\footnote{\url{https://ec.europa.eu/commission/presscorner/detail/en/IP_23_2413}}
Nonetheless, it remains uncertain whether researchers can obtain the necessary data to expand on this research.

The heuristics and detection methods introduced in our study do not necessarily work on other generative AI models.
For example, diffusion models~\citep{song2019generative,ho2020denoising} yield better image-generation outcomes than GANs~\citep{dhariwal2021diffusion,yang2023diffusion}.
This architecture has been adopted by many AI companies, including OpenAI~\citep{ramesh2022hierarchical}, Midjourney,\footnote{\url{https://docs.midjourney.com}} and Stability AI,\footnote{\url{https://stability.ai/stable-image}} to create advanced multimodal models.
The images generated by these models are astonishing and free from the issues of GAN-generated ones.
Preliminary studies on detecting diffusion model-generated images suggest that the techniques that work on GAN-generated images are no longer effective~\citep{ricker2022towards}, requiring new detection paradigms.

Despite these limitations, our findings provide empirical evidence of the threats posed by generative AI.
We show that GAN-generated images have already been deployed at scale to create large amounts of convincing fake accounts.
Many more social media accounts likely have already been using more advanced AI models to generate more realistic fake profiles.
We even uncover a group of accounts that also leverage ChatGPT to generate human-like text and conduct realistic interactions~\citep{yang2024anatomy}.
Combining these AI models can lead to more convincing and realistic fake accounts on social media. 
In fact, recent research suggests that advanced AI models enable the creation of autonomous agents that can ingest multimodal inputs, process the data, reason, and then leverage various tools to achieve their objectives~\citep{xi2023rise}.
Despite the usefulness of these intelligent agents, the risk of adversarial actors abusing them is high.

The looming threats call for new intervention strategies.
First, we need effective models that can detect AI-generated images.
A recent study finds that assistance from such models can improve people's performance in identifying GAN-generated images~\citep{boyd2023value}.
However, the research community might struggle to catch up as these cutting-edge generative models evolve quickly.
Another approach requires regulations on the use of generative AI models to create inauthentic accounts and content. 
For instance, social media platforms can require an account to provide evidence of their authenticity before exposing them and their content to a broader population~\citep{Menczer2023AI-harms}. 
Last but not least, we believe that teaching social media users AI literacy is crucial.
Understanding the capability and weaknesses of state-of-the-art generative AI models can help social media users better defend themselves from manipulation by AI-powered fake accounts. 
However, such initiatives should be carried out carefully to avoid unintended consequences~\citep{yan2022exposure}. 

\section*{Acknowledgements}

We thank \url{https://generated.photos} for sharing their datasets for our research and Twitter user @conspirator0 for flagging candidate accounts using GAN-generated profile images. 

\bibliography{ref}

\end{document}